\documentclass[preprint,showpacs,preprintnumbers,amsmath,amssymb,superscriptaddress]{revtex4}

\usepackage{graphicx}
\usepackage{dcolumn}
\usepackage{bm}

\usepackage{xcolor}
\definecolor{granata}{HTML}{831d1c}
\definecolor{kulblue}{HTML}{116E8A}

 \newcommand{\bfE}{\mathbf{E}}

\newcommand{\bfB}{\mathbf{B}}

\newcommand{\bfS}{\mathbf{S}}

\newcommand{\bfJ}{\mathbf{J}}

\newcommand{\bfV}{\mathbf{V}}

\begin{document}

\title{Topographic analysis of  fluctuations in  3D reconnection. }%
\title{A violin sonata for reconnection. }
\author{G. Lapenta} 
\affiliation{Department of Mathematics, Center for mathematical Plasma Astrophysics, KU Leuven, University of Leuven, Belgium}
\author{F. Pucci} 
\affiliation{Department of Mathematics, Center for mathematical Plasma Astrophysics, KU Leuven, University of Leuven, Belgium}
\author{M.V. Goldman } 
\affiliation{Department of Physics, University of Colorado, Boulder, USA}
\author{D.L. Newman} 
\affiliation{Department of Physics, University of Colorado, Boulder, USA}

\date{\today}

\begin{abstract}
The process of magnetic reconnection when studied in Nature or when modeled in 3D simulations differs in one key way from the standard 2D paradigmatic cartoon: it is accompanied by much fluctuations in the electromagnetic fields and plasma properties. We developed a new diagnostics, the \textit{topographical fluctuations analysis (TFA)} to study  the spectrum of fluctuations in the various regions around a reconnection site. We find that fluctuations belong to two very different regimes. The first regime is better known, it develops in the reconnection outflows and is characterized by a strong link between plasma and electromagnetic fluctuations leading to  momentum and energy  exchanges via anomalous viscosity and resistivity. But there is a second, new, regime: it develops in the inflow and in the region around the separatrix surfaces, including the reconnection diffusion region itself. In this new regime the plasma remains laminar but the electromagnetic fields fluctuates strongly. We present an analogy with the smooth continuous motion of the bow of a violin producing the vibrations of the strings to emit music. \end{abstract}

\maketitle

%
%

%


%
%
%
%

The cartoon image of reconnection is laminar. Reconnection is thought as two flied lines coming together from regions of opposite magnetic polarity, breaking and reconnecting to form two new field lines with different topological connection \cite{biskamp}. The process is associated with considerable magnetic energy release into kinetic particle energy. This picture naturally associates great value into the central point where the field lines touch and reconnect, the x-point as it is called. In 2D thinking this model is extremely powerful and can explain much if not most of what is observed.  The magnetospheric multiscale (MMS) mission confirmed or uncovered many features of reconnection that can be understood within this paradigm~\cite{burch2016electron}.  

However, it has long been known that reconnection is also associated with waves and fluctuations~\cite{ji2004electromagnetic}. MMS has reconfirmed this point \cite{ergun2016magnetospheric,ergun2017drift}. 
In recent studies focused on  dayside reconnection, where MMS spent the first phase of its mission, the presence of fluctuations has also been reproduced in full kinetic particle simulations~\cite{price2017turbulence}. We look here at the conditions expected for fluctuations in the nightside of the Earth magnetosphere, the magnetotail.  

The conditions are different because the magnetotail has  symmetric density, magnetic field (and plasma beta). Additionally the dayside can have a strong field in the out of plane direction (called guide field and directed in the dawn-dusk direction along the Earth rotation) while in the tail such field is usually rather weak. 

In the present letter, we consider the different regimes of fluctuation present in different regions around a reconnection site for conditions typical of the magnetotail. To reach this goal we developed a new investigation method, the \textit{topographical fluctuations analysis (TFA)} designed to provide statistical information on the fluctuation spectra in different regions. The outcome of the study is perhaps surprising and certainly at considerable variance with the results obtained in the magnetopause~\cite{price2017turbulence}. We identify two regimes of fluctuations. One in the outflow leads to a turbulent regime  where the fluctuations involve both fields and particles. In the inflow and separatrix region, instead, the fluctuations involve only the fields without affecting the particles. 

The two regimes differ much in practical consequences. The outflow MHD regime is capable of inducing a strong and turbulent energy exchange as well as strong anomalous momentum exchange dominated primarily by the electrostatic term in Ohm's law. The inflow regime, instead, does not lead to substantial fluctuations in the energy exchange nor significant momentum anomalous dissipations. This is very different from the magnetopause where fluctuations-induced anomalous effects are strongly present also in the separatrix and even electron diffusion region~\cite{price2017turbulence}.

We present a visual analogy to understand this regime. We compare the electromagnetic fluctuations with the strings of a violin and the plasma motion around a reconnection site as a bow of a violin: the strings vibrate under the continuous friction of the bow. The inflow and separatrix fluctuations are similar: the electromagnetic fields oscillate under a continuously streaming plasma. But the streaming plasma does not vibrate, just like the bow of a violin moves smoothly in the expert hands of a violinist.

\begin{figure}[htbp]
\includegraphics[width=.9\columnwidth]{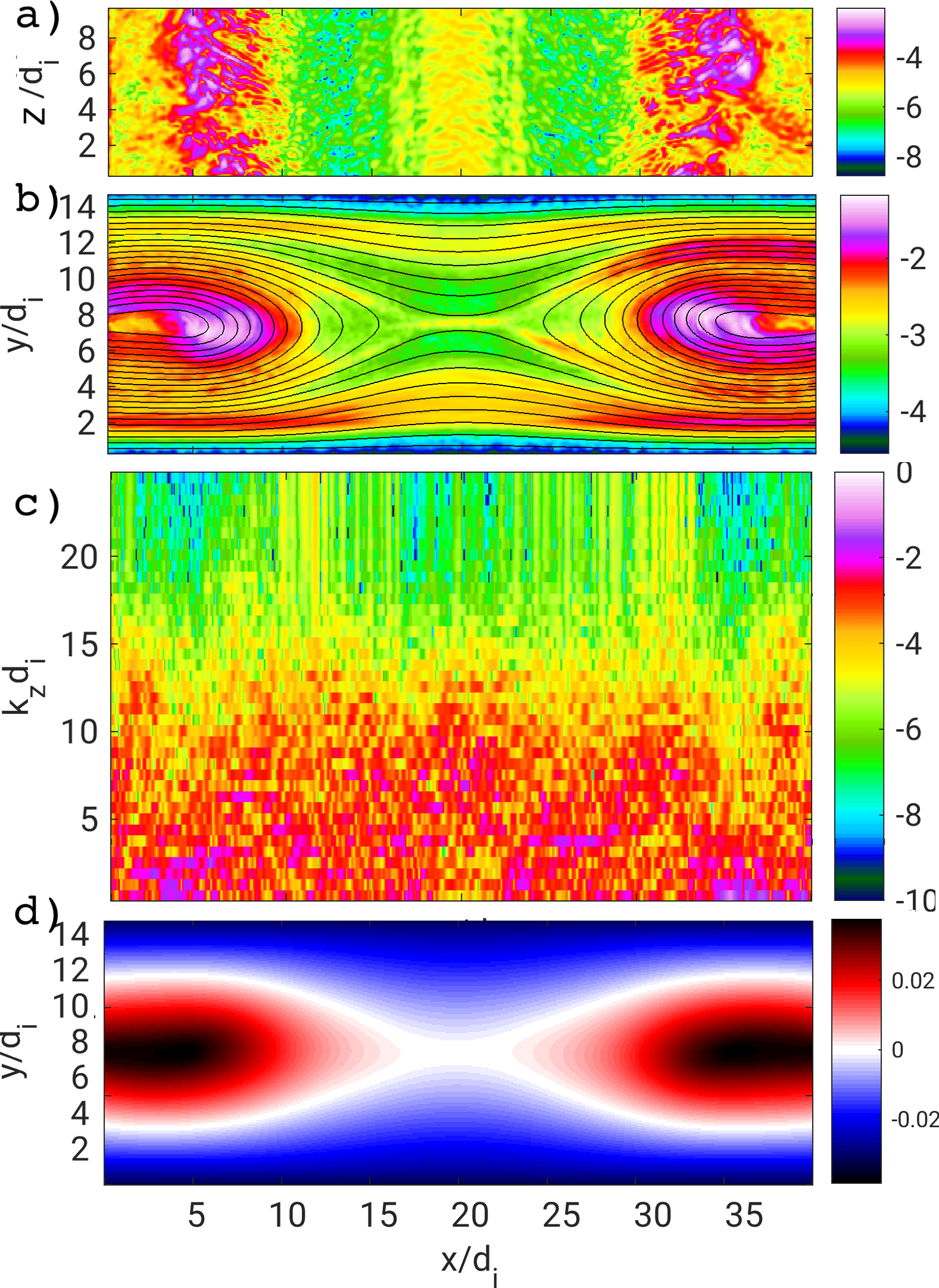}
\caption{Divergence of the Poynting flux at $\omega_{ct}=21.825$.  From top to bottom are reported: a) false color representation of $c\nabla \cdot \bfS/m_i\omega_{pi}^4$ in the mid-plane $y=L_y/2$; b)  the standard deviation of $\nabla \cdot \bfS$  along $z$, superimposed with the contour plot of the mean flux function $\Psi$;  c) spectrum of Fourier modes  $\widetilde{\nabla \cdot \bfS}(x,y,k_z)$  along the $x$ axis (at $y=L_y/2$), normalized to their integral $\int dk_z \widetilde{\nabla \cdot \bfS}(x,y,k_z)$; d) flux function $e\Psi(x,y)/m_ic$ averaged over $z$. }
\label{figure1}
\end{figure}
Our investigation uses a 3D fully kinetic particle in cell approach based on the code iPic3D \cite{markidis2010multi}. The present simulation has a mass ratio $m_i/m_e=256$, $v_{the}/c=0.045$, $T_i/T_e=5$. The initial condition is that of a Harris sheet with thickness $L/d_i=0.5$ in a domain where the initial magnetic field is along $x$ with size  $L_x=40d_i$, the initial gradients are along $y$ with $L_y=15d_i$. The third dimension, where the initial current is directed,  is initially invariant with $L_z=10d_i$. Open boundaries are imposed in $x$ and $y$ and periodicity is assumed along $z$.

The Harris plasma density peaks in the central plane $y=L_y/2$ and decays exponentially to zero along $y$. The real magnetotail has a non zero asymptotic density and to increase the level of fidelity we also add a background of 10\% of the peak Harris density. We choose to set the simulation initially  in the ion rest frame so that all the current is initially carried by the electrons. In this frame then the background plasma is not drifting and the overall system has no velocity shear (since the electron mass is much smaller and the center of mass speed is essentially the speed of the ions that is initially zero). This choice avoids the presence of shear-driven modes that tend to kink the current sheet, an effect that would have complicated the interpretation of the fluctuations \cite{lapenta2003unexpected}. The four plasma species (ions and electrons in the Harris and background) are each described by 125 particles per cell (with non uniform weight in the case of the Harris species). 
The grid  has 512x192x128 cells, resolving well the electron skin depth, in the reconnecting background plasma $\Delta  x \approx 0.4 d_{e,b}$ (the initial Harris plasma is quickly swept away by reconnection). The time step also resolves equally well the electron cyclotron frequency, even in the strongest field, $\omega_{ce,Bmax}\Delta t \approx 0.3$ (and even better in more average fields).

Figure \ref{figure1} shows the standard deviation of the fluctuations  of the divergence of the Poynting flux, $\nabla \cdot \bfS$. As can be observed, fluctuations are present everywhere, but focusing on the regions  exceeding the -5 decade in logarithmic scale (yellow to white), we can distinguish three regions: 1) the inflow region above and below the central reconnection region, 2) the central reconnection region, with a band around the separatrices, 3) the outflow region where the downstream pileup front form.

A cut through the mid plane (Fig. \ref{figure1}-a) highlights the structure of these fluctuations.  A quasi-periodic repeating pattern is evident but a study based on 1D Fourier decomposition along $z$ (see Fig. \ref{figure1}-c) reveals a turbulence-like non linear cascade of Fourier modes along $z$ \cite{pucci2017properties}. The spectrum in  $k_{z}$ is rather similar in all regions at different distances from the reconnection site including the electron diffusion region itself. Note that the spectrum is normalized to its integral for each $x$, the fluctuation levels are much weaker in the central reconnection region than in the outflow but the spectrum in $k_z$ is rather invariant, denoting an independence to the original mechanism producing the fluctuations and the establishment of a non-linear turbulent cascade.

The spatial invariance of the spectrum of $k_{z}$ might be indicative of a common cause. The fluctuations around the reconnection region are caused by the drift of the species, different for electrons and ions. In the inflow along the separatrices (see Fig.~\ref{figure3}-b), the electrons drift causes instabilities due to the relative ion-electron drift (Buneman instability~\cite{Goldman2008,Divin2012}) and between different electron populations~\cite{Goldman2014}. In the outflow, the dominant instability is the lower hybrid drift instability due to the strong density gradients formed in the pile up region where the outflow meets the ambient plasma \cite{divin2015evolution,lapenta2018nonlinear} and the unfavorable curvature of the field lines contributes to driving the instability \cite{guzdar2010simple}. Once the fluctuations are generated a process of nonlinear cascade ensues leading to a turbulent behavior \cite{vapirev2013formation,lapenta2015secondary,innocenti2016study}
To analyze the fluctuations in each region, we have developed a specific diagnostic referred to as \textit{topographic fluctuation analysis (TFA)}. The idea is to use the same statistical tools used in demographics. A statistician might pose the question of what is the income distribution among the people living in different districts of a city. Similarly we study the spectrum of  fluctuations in different regions around the reconnection region.

The main question is how do we identify the regions? The flux function is a natural choice, considering that the simulation is initially invariant in $z$ and throughout the evolution this invariance remains valid on average, except for the fluctuations. We define in each plane at constant $z$ a flux function $\Psi(x,y)$, defined as the scalar function whose contours are everywhere parallel to the magnetic field on that plane (i.e. the function is defined by the generating equation $\nabla_{xy} \Psi \cdot \bfB=0)$. Note that this function is not the out of plane component of the vector potential: this property will be true only in a 2D domain. In our 3D case, the flux function defined above is just a useful function to define the intersection of the magnetic field surfaces with a plane at given $z$. 

The flux function defined above can be shifted in each plane by a constant: we define it so that the flux function in each plane is  zero at the intersection where the separatrix surfaces meet $\Psi(x_{x},y_{x})=0$ (where $x_{x},y_{x}$ identifies the  so-called x-point in each plane). Figure 1-d reports an example of such flux function for the central plane ($z=Lz/2$). It obviously resembles the out of plane component of the vector potential in 2D domains despite the warning above. 

To define the  TFA, we subdivide the range of  $\Psi$ into 100 3D regions between two surfaces at two consecutive values of $\Psi$. The regions are 3D because $\Psi$ is defined in every $z$ plane. The region  around $\Psi=0$, by construction, is centered on the reconnection site.

We then measure in each of these regions how the fluctuations for a given quantity are statistically distributed.  In this task, we use the convenient periodicity along $z$ where all the quantities are nearly invariant except for the fluctuations. A fluctuation can then be simply defined as the variance with respect to the value averaged along $z$ at a given $(x,y)$ location. With this definition of fluctuations, we bin the fluctuations in each region into 100 bins measuring the  occurrences of the values of the fluctuations. In essence we count how many times a fluctuating quantity is in between two consecutive values in a range of 100 intervals between maximum and minimum. The maximum and minimum are defined over the whole domain. This results in an occurrence count that is different for each region of $\Psi$. 
 
The approach is based on the demographic analogy. What we do in essence is bin all people into 100 income groups from the richest to the poorest person in the city, and measure the number for each income bin in each city district. The result is a 2D map where for each district we know the count of incomes.  In our study we obtain the count of fluctuation levels in 100 different fluctuation bins for each of the 100 flux function regions in which the domain is divided.

\begin{figure}[htbp]
\centering
\includegraphics[width=.9\columnwidth]{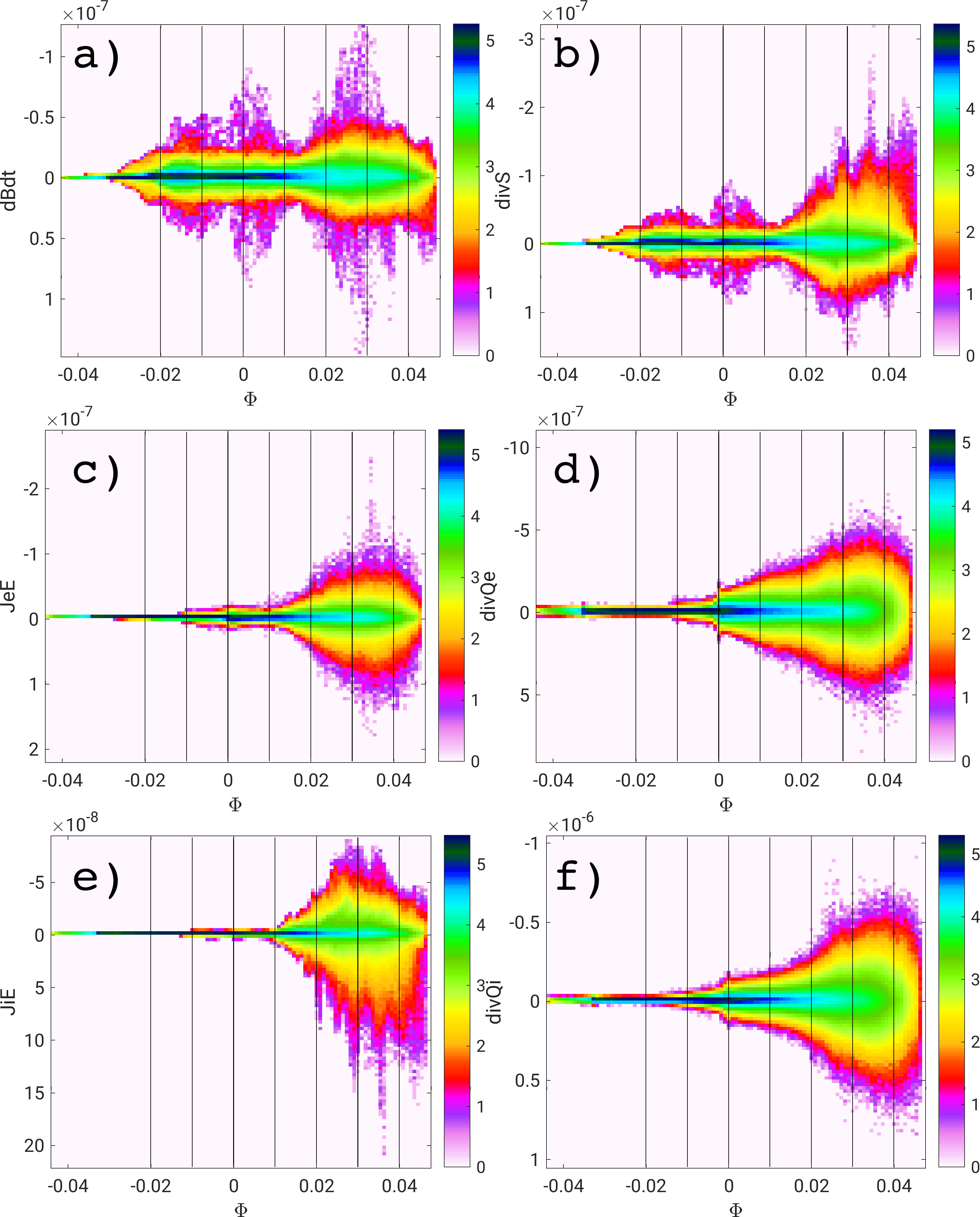}
\caption{Results of the TFA analysis displaying the occurrence count in 100 fluctuation ranges for 100 flux function domains. The quantities measured from top to bottom and left to right are: a) magnetic energy ($W_B=e^2B^2/2\mu_0 m_i^2 \omega_{pi}^2$) change $\omega_{pi}^{-1}\partial W_B/\partial t$, b) divergence of the Poynting flux $c\nabla \cdot \bfS/m_i \omega_{pi}^4$, c) $c\bfJ_e \cdot \bfE /m_i \omega_{pi}^4$, d) divergence of the electron energy flux, e) $c\bfJ_i \cdot \bfE /m_i \omega_{pi}^4$, d) divergence of the ion energy flux.}
\label{figure2}
\end{figure}

The resulting analysis is shown in Fig. \ref{figure2}. What can be observed is the fluctuation range, defined with a sign because fluctuations can be positive or negative relative to the mean. In each panel we can observe the fluctuation spectrum in the region where the separatrices meet (around  $\Psi=0$), in the inflow ( $\Psi<0$) and in the outflow ( $\Psi>0$). The precise location corresponding to a given value of $\Psi$ can be obtained form Fig.~\ref{figure1}-d. 

Each row in Fig.~\ref{figure2} corresponds to an energy channel. From the top, we report first the electromagnetic energy exchange between Poynting flux and magnetic energy (the electric field energy is negligible in our simulation where the speeds remain much below the speed of light). Next we report the electron energy exchange with the electromagnetic field and the electron particle energy flux (inclusive of all energy fluxes: bulk, enthalpy and heat flux). Finally the ions are shown. 

The most surprising feature is that in the inflow ($\Psi<0$) and reconnection region proper ($\Psi\approx 0$) almost only the electromagnetic energy is being exchanged by the fluctuations. In essence the magnetic energy is converted to Poynting flux and viceversa. The fluctuations affect the electrons and ions minimally, except  far downstream of the reconnection region for $\Psi>0$. The  energy exchange rate with the electric field, $\bfJ_s \cdot \bfE$,  fluctuates only in the far outflow, while the species energy flux  starts to fluctuate  at the central  reconnection site at $\Psi \geq 0$. 


The TFA clearly identifies two different regimes for the fluctuations. The first is in the region of inflow  ($\Psi<0$) where only the electromagnetic field fluctuate but the particle energy exchange with the fields ($\bfJ_{s} \cdot \bfE$) does not respond to the fluctuations. The second region is that of the outflow ($\Psi>0$) where the species start to respond to the fluctuations. 

\begin{figure}[htbp]
\includegraphics[width=\columnwidth]{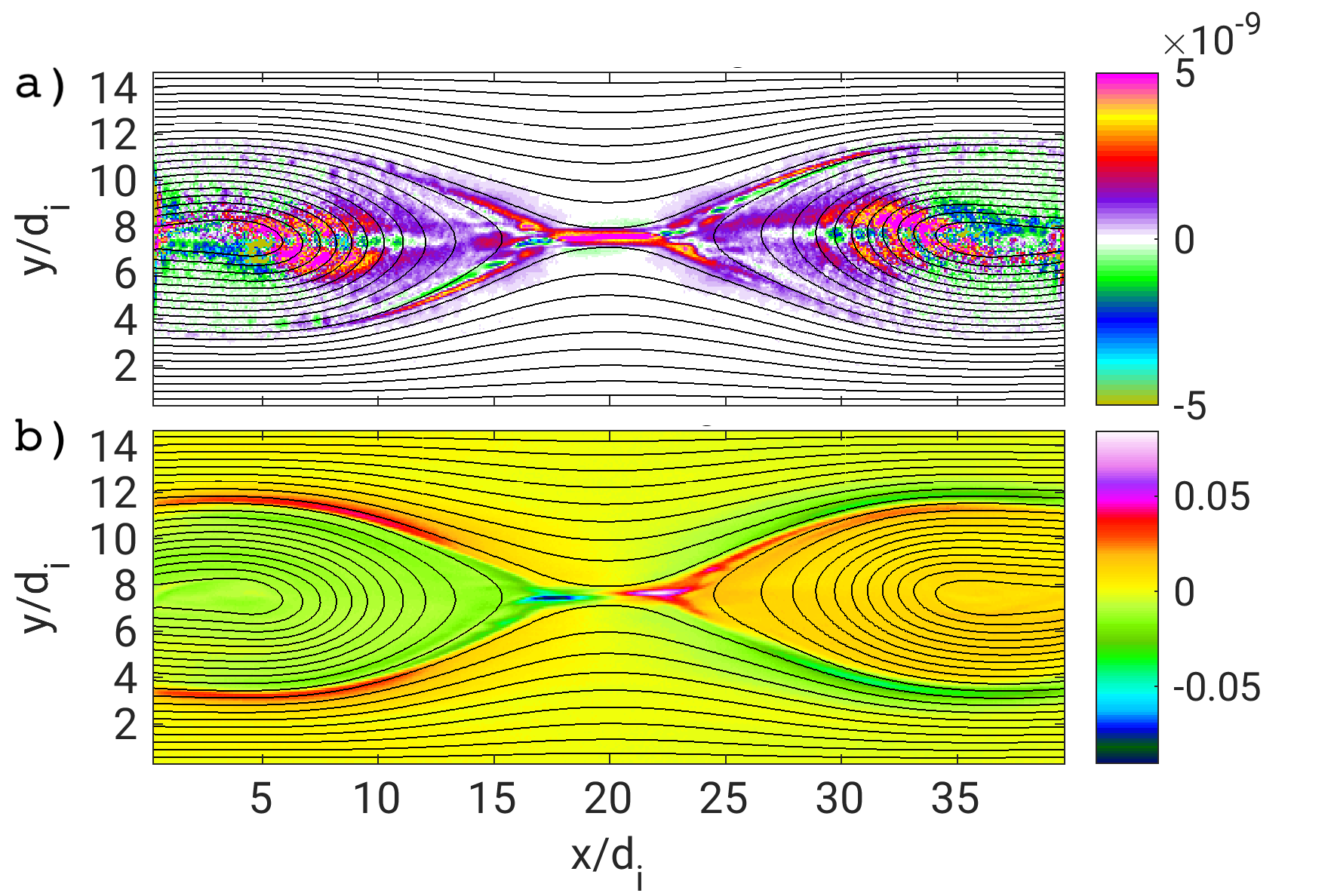}
\caption{False color representation of the electron energy exchange rate with the electromagnetic fields, $c\bfJ_e \cdot \bfE /m_i \omega_{pi}^4$ (a, top) and of the electron horizontal speed, $V_{ex}/c$ in the mid plane $z=L_z/2$ (b, bottom). Average along the $z$ direction.}
\label{figure3}
\end{figure}

These two regimes of fluctuation identify two different types of processes. In the reconnection region proper, the region near the separatrices and the inflow the instabilities and their non-linear development do not cause particle fluctuations. But the reader should not jump to the conclusion that there is no energy exchange. Figure \ref{figure3} shows the  electron bulk flow and the electron energy exchange rate with the electric field. The energy exchange is just as large in the central and separatrix region as it is in the outflow, but it remains laminar. 

We propose a musical analogy: as a violin player slides the bow over a string, the string vibrates at high frequency emitting the music but the bow moves regularly and smoothly, transferring its energy to the string monotonically. Over each swipe of the bow the string vibrates many times, but the bow moves steadily at constant speed. What we see in this violin-like fluctuation regime is that the electromagnetic fields fluctuate but the particles act as a laminar flow. The fluctuations do not carry over to the particles: the waves act in an average way and the particles gain or loose energy in an average sense.

The  regime in the outflow is more well known and  is characterized by  fields and plasma species fluctuating together. This process develops fully only in the pile up regions at the end of the outflows. Previous studies already identified a developing turbulence in this region and the reader is referred to those for more details \cite{lapenta2016energy,pucci2017properties,pucci2018generation,lapenta2018nonlinear}. 

A key consequence of these two distinctly different regimes of fluctuation is that the violin-like regime should be incapable of really affecting the momentum equation by producing an anomalous term because it is incapable of forcing fluctuations in the plasma species. Conversely, the fluctuation regime in the outflow could lead to strong anomalous effects.

The study of anomalous momentum exchange rates is based on the generalized Ohm's law and on considering for each quantity an average value and the fluctuations. For the electric field for example, $\bfE= \langle\bfE\rangle +\delta \bfE$. Substituting the decompositions for each quantity into the generalized Ohm law and  averaging it along $z$, one obtains \cite{braginskii}:
\begin{equation}
e \langle n \rangle \langle \bfE \rangle = m_e \langle n\rangle \frac{d\langle\bfV_e\rangle}{dt}
+ \nabla \cdot  \langle P_e \rangle -\langle \bfJ_e \rangle \times \langle \bfB \rangle  +\Xi,
\end{equation}
where the anomalous terms $\Xi$ are:
\begin{equation}
\Xi = - e \langle \delta n \delta \bfE \rangle + m_e \langle \delta n \frac{d\delta\bfV_e}{dt} \rangle -\langle \delta \bfJ_e  \times \delta \bfB \rangle  ,
\end{equation}
where the three fluctuation-supported terms are the electrostatic (aka anomalous resistivity), inertia and electromagnetic (aka anomalous viscosity~\cite{price2017turbulence}) terms. The fluctuations of the pressure tensor in this formalism do not produce anomalous effects.

\begin{figure}[htbp]
\includegraphics[width=\columnwidth]{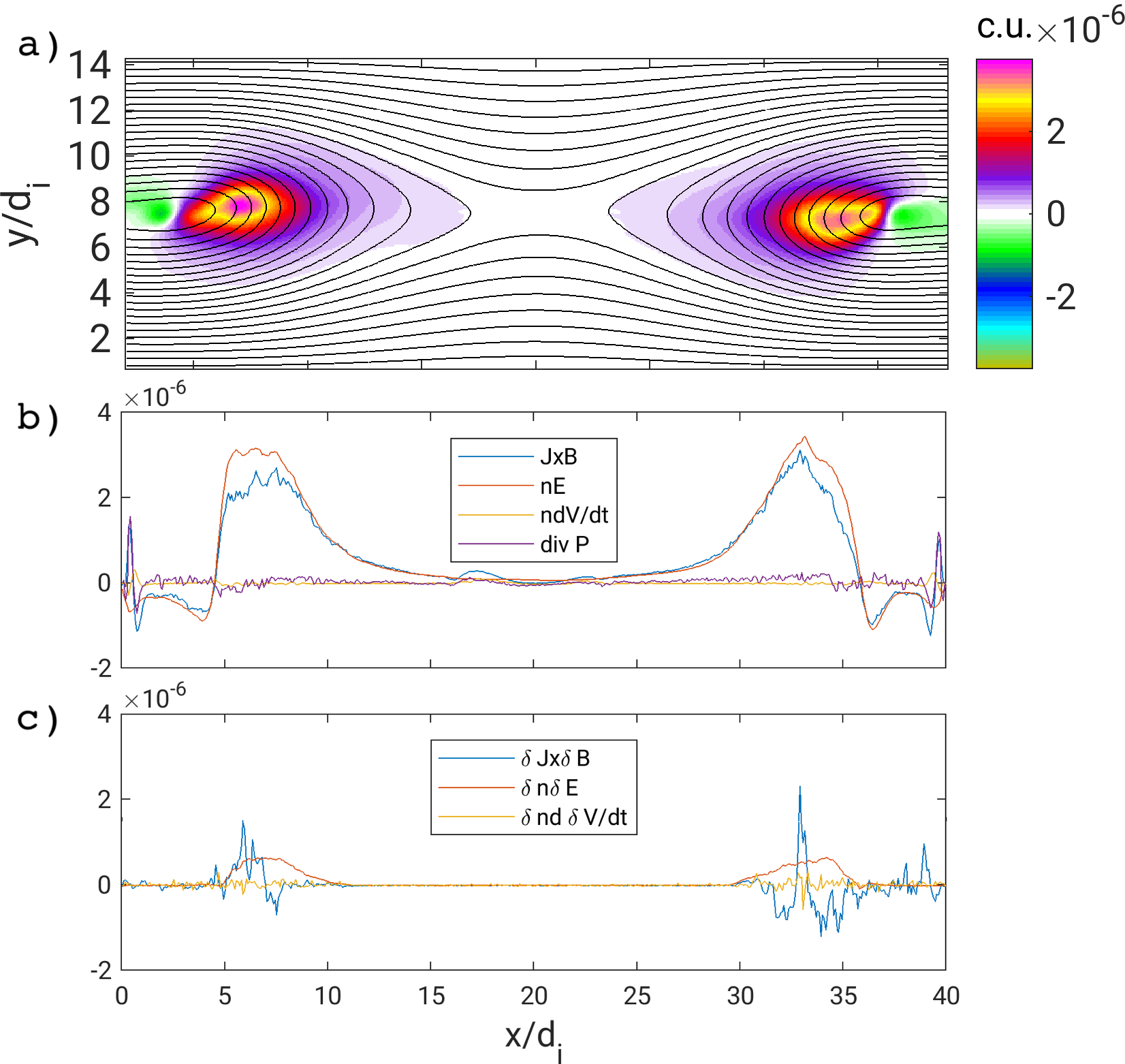}
\caption{False color representation of the right-hand side of the generalized Ohm's law, $e\langle n \rangle \langle \bfE \rangle c^2 /m_i \omega_{pi}^4$ (top). Cut along the $x$ axis (at $y=L_y/2$) of the mean  (middle)  and fluctuation (bottom) terms of  the generalized Ohm's law.}
\label{figure4}
\end{figure}

Figure \ref{figure4} shows the different terms in the generalized Ohm's law. As predicted above, the region of violin-like fluctuations are not providing any momentum exchange, the mean and the fluctuation terms remain negligible. 

Contrary to what one could think, the action in reconnection is not at the center: the electron diffusion region at this stage of reconnection is not a center of intense momentum or energy exchange. However, this is not the case at the early stages of reconnection onset, then all the action is in the center.
At this advanced state when reconnection is fully established, all the action is on the outflow fronts.  In Fig.~\ref{figure4}, the top panel shows the main term $\langle n\rangle\langle E\rangle$. The second panel reports the main terms of the averaged generalized Ohm's law. The last panel reports the so-called anomalous contributions to the generalized Ohm's law produced by the fluctuations. 

The majority of the out of plane electric field is caused by the Hall and advection  term $\langle J_e\rangle \times \langle \bfB\rangle$, with the pressure term being small in comparison. The anomalous terms, instead, provide a substantial minority contribution, especially the electrostatic fluctuation term, $\delta n \delta E$ but also the electromagnetic term, $\delta J_{e} \times \delta \bfB$. The inertia term is negligible by virtue of the mass ratio of the electrons being very small in the present simulation, a realistic effect because in reality it is even smaller (1836 instead of 256 for hydrogen plasma).

In summary, this letter analyses the fluctuations around a reconnection site using 3D PIC simulations with electrons and ions both treated as  particles. The resolution includes the electron effects down to a fraction of the electron skin depth and electron cyclotron frequency, well beyond the scale of interest of the fluctuations studied. The results can then be considered well converged, as proven by simulations with less resolution but substantially equivalent results not reported here. The main conclusion is that fluctuations in the outflow present the classic scenario of turbulent reconnection with plasma species and electromagnetic fields fluctuating in sync. The inflow and separatrix region, instead, present a different type of fluctuations: the fields fluctuate but the plasma species remain essentially laminar. These electromagnetic fluctuations are compared by analogy with the vibration of the string of a violin in response to the monotonic motion of the bow: electron and ions exchange energy with the fields but retain their laminar motion. 

The consequence of these two regimes is that the fluctuations produce strong momentum and energy exchange with the fluctuations only in the outflow resulting in large energization of the plasma species and large anomalous dissipations (viscosity and resistivity) only in the outflow. In the inflow and separatrix regions there is strong but laminar energy  exchange but essentially no anomalous momentum exchange despite the electromagnetic fluctuations present. 

The conclusions reported here are valid for conditions in the Earth nightside reconnection  and differ starkly with those reported for the Earth dayside reconnection \cite{price2017turbulence}. The data from the MMS mission provides direct in situ opportunity to hunt for the new violin-like turbulent state reported here: finding proof of its actual existence  is the natural follow up of our work.

{\it 
This project has received funding from the European Union's Horizon 2020 research and innovation programme under grant agreement No. 776262 (AIDA). This
research used resources of the National Energy Research
Scientific Computing Center, which is supported by the Office
of Science of the US Department of Energy under Contract No.
DE-AC02-05CH11231. Additional computing has been provided
by NASA NAS and NCCS High Performance Computing,
by the Flemish Supercomputing Center (VSC) and by a
PRACE Tier-0 allocation.}

\bibliography{lapenta}
\end{document}